\documentstyle[aas2pp4]{article}

\begin{document}

\title{Orbital Parameters for the Soft X-ray Transient 4U 1543-47:
Evidence for a Black Hole}

\author{Jerome A. Orosz\altaffilmark{1}}
\affil{Department of Astronomy \& Astrophysics, The Pennsylvania State
University, 525 Davey Laboratory, University Park, PA  16802-6305 \\
orosz@astro.psu.edu}

\author{Raj K. Jain\altaffilmark{1,2} and Charles D. Bailyn\altaffilmark{3}}
\affil{Department of Astronomy, Yale University, P.O. Box 208101,
New Haven, CT 06520-8101  \\ 
raj.jain@yale.edu, bailyn@astro.yale.edu}

\author{Jeffrey E. McClintock}
\affil{Harvard-Smithsonian Center for Astrophysics, 60 Garden Street,
Cambridge,  MA  02138-1516 \\
jem@cfa.harvard.edu}

\and

\author{Ronald A. Remillard}
\affil{Center for Space Research, Massachusetts Institute of Technology,
Cambridge, MA 02139-4307 \\ rr@space.mit.edu}

\altaffiltext{1}{Visiting Astronomer at Cerro Tololo Inter-American
Observatory (CTIO), which is operated by the Association of
Universities for Research in Astronomy Inc., under contract with the
National Science Foundation.}
\altaffiltext{2}{Also Department of Physics, Yale University}
\altaffiltext{3}{National Young Investigator}

\begin{abstract}
Spectroscopic observations of the soft X-ray transient 4U 1543-47
reveal a radial velocity curve with a period of $P=1.123\pm 0.008$
days and a semi-amplitude of $K_2=124\pm 4$~km~s$^{-1}$.  The
resulting mass function is $f(M)=0.22\pm 0.02\,M_{\sun}$.  We classify
the secondary star as A2V, in agreement with previous work, and
measure $T_{\rm eff}= 9000\pm 500$~K and $E(B-V)=0.50\pm 0.05$ from
fits to synthetic spectra.  We derive a distance of $d=9.1\pm 1.1$~kpc
if the secondary is on the main sequence.  We see no emission lines
from the accretion disk, and the measured fraction of disk light in
the $B$ and $V$ bands is 10\% and 21\% respectively.  The $V$ and $I$
light curves exhibit two waves per orbital cycle with amplitudes
$\approx 0.08$ mag.  We modeled the light curves as ellipsoidal
variations in the secondary star and derive extreme inclination limits
of $20\le i\le 40\arcdeg$.  The formal $3\sigma$ limits on the
inclination from a simultaneous fit to the $V$ and $I$ light curves
are $24\le i\le 36\arcdeg$ for a mass ratio $Q\equiv M_1/M_2\ge 1$.
However, there are systematic effects in the data that the model does
not account for, so the above constraints should be treated with
caution.  We argue that the secondary star is still on the main
sequence with the mass transfer being driven by expansion due to
normal main sequence evolution.  If the secondary star has a mass near
the main sequence values for early A-stars ($2.3\le M_2\le
2.6\,M_{\sun}$), then the best fits for the $3\sigma$ inclination
range ($24\le i\le 36\arcdeg$) and the $3\sigma$ mass function range
($0.16\le f(M)\le 0.28\,M_{\sun}$) imply a primary mass in the range
$2.7\le M_1\le 7.5\,M_{\sun}$.  The mass of the compact object in 4U
1543-47 is likely to be in excess of $\approx 3\,M_{\sun}$, which is
widely regarded as the maximum mass of a stable neutron star.  Thus we
conclude 4U 1543-47 most likely contains a black hole.
\end{abstract}


\section{Introduction}\label{intro}

4U 1543-47 is a bright soft X-ray transient that was observed in
outburst in 1971 (Matilsky et al.\ 1972\markcite{mat72}; Li et al.\
1976\markcite{li76}), 1983 (Kitamoto et al.\ 1984\markcite{kit84}),
and 1992 (Harmon et al.\ 1992\markcite{har92}).  This object is
considered by many to be a strong black hole candidate based on its
X-ray spectrum, which is composed of an ultrasoft component and a hard
power-law tail (Tanaka \& Lewin 1995\markcite{tl95}).

The optical counterpart (IL Lup) was discovered shortly after the 1983
outburst (Pederson 1983\markcite{Ped83}).  In quiescence the
counterpart turned out to be unusual in several ways.  For example
Chevalier (1989)\markcite{chev89} reported that the counterpart had
faded by only 1.8 mag in $V$, much less than the $\approx 7$ mag
typical for soft X-ray transients (van Paradijs \& McClintock
1995\markcite{vPMc95}).  Also, the spectral type of the secondary was
A2V, much earlier than the usual K-type stars found in systems like
A0620-00 or Nova Mus 1991.  Finally, Chevalier noted that the spectrum
of the quiescent counterpart lacked emission lines of any kind,
whereas one expects to see strong Balmer emission lines from an
accretion disk (van Paradijs \& McClintock 1995\markcite{vPMc95}).
Chevalier did not state whether any photometric variability or radial
velocity variations were observed.  He proposed three hypotheses to
explain the A2V star at the position of 4U 1543-47: ~~a)~The A star is
in the line of sight to the X-ray source and not physically associated
with it; ~~b)~The A star is a very ``unusual'' mass donor star;
~~c)~The A star is the outer star of a triple system with a low-mass
X-ray binary as the inner pair.

GRO J1655-40, an important X-ray transient, was discovered in 1994
(Zhang et al.\ 1994\markcite{zh94}).  The optical counterpart of this
source (V1033 Sco) has some characteristics similar to those of the
counterpart of 4U 1543-47. First, the optical outburst amplitude was a
modest 3.3 mag in $V$ (Bailyn et al.\ 1995a\markcite{bail95a}).  In
addition, the secondary is an early type star (F5 IV, Bailyn et al.\
1995b\markcite{bomr95b}; Orosz \& Bailyn 1997a, hereafter
OB97\markcite{ob97}).  Finally, the optical spectrum of GRO J1655-40
in X-ray quiescence is devoid of emission lines (OB97\markcite{ob97}).
Motivated by these similarities, we observed the counterpart of 4U
1543-47 and discovered that the second hypothesis of Chevalier
(1989)\markcite{chev89} is the correct one---the A star is in fact the
mass donor, although in view of GRO J1655-40 with its F-star
secondary, the A star is perhaps not an ``unusual'' mass donor.  We
describe below our observations and reductions, data analysis, and
conclusions.

\section{Observations and Reductions}\label{obs}

Images of 4U 1543-47 were obtained 1997 June 28-July 3 with the CTIO
1.5 m telescope, the Tek $1024\times 1024$ \#2 CCD at the f/7.5 focal
position, and standard $B$, $V$, and $I$ filters.  The nights were
generally clear during the time 4U 1543-47 was visible, and the seeing
ranged from $1\farcs 5$ to $2\farcs 2$.  Standard IRAF tasks were used
to process the images to remove the electronic bias and to perform the
flat-field corrections.  Observations of several standard star fields
from Landolt (1992)\markcite{landolt92} taken June 28 and June 29 were
used to calibrate the photometry to the standard system.  The
calibrations from the two nights agreed to within 0.02 magnitudes.

Spectra of 4U 1543-47 were taken 1997 July 1-4 at the V. M. Blanco 4 m
telescope at CTIO using the R-C spectrograph and the Blue Air Schmidt
camera + Loral $3072\times 1024$ CCD.  To reduce the readout noise,
the CCD was binned on chip by 2 in the spatial direction.  The KPGL
\#3 grating (527 $\ell$/mm, first order) and the blue collimator were
used on the night of July 1.  The resulting ``blue spectra'' cover the
wavelength range 3466-7162~\AA\ in 1.2~\AA\ pixels; the spectral
resolution is about 3.6~\AA\ FWHM.  The G420 grating ($600\ell$/mm,
first order) and the red collimator were used the nights of July 2-4
to obtain ``red spectra'' that cover 5820-9130~\AA\ with 1.08~\AA\
pixels and have a spectral resolution of about 3.3~\AA\ FWHM.  The
seeing was typically 1\farcs 0 to 1\farcs 5, although on some
occasions it reached 2\farcs 0.  The nights of July 1-3 were
photometric during the observations of 4U 1543-47.  However on July 4
there was 1 to 2 magnitudes of extinction due to clouds.  The slit
width was fixed at 1\farcs 5 for all observations, and the slit angle
was rotated to match the parallactic angle (Filippenko
1982\markcite{fil82}) for all of the July 1-3 observations.  The slit
angle on July 4 was set to $90\arcdeg$ (east-west) for the bright
comparison stars and $75\arcdeg$ for 4U 1543-47.  Directly after the
telescope was moved or the spectrograph was rotated, a He-Ne-Ar lamp
was observed.  During extended observations of 4U 1543-47, the lamp
was observed every hour.  The dispersion solutions are accurate to
better than 0.1~\AA, and the maximum spectrograph flexure that
occurred during an entire night was less than 2 pixels.  We obtained a
total 41 of spectra of 4U 1543-47 over the 4 nights with exposure
times of 900-1200 seconds. We also obtained 28 spectra of 12 different
A-type comparison stars, 13 spectra of 12 IAU radial velocity standard
stars, 10 spectra of 4 different flux standard stars, and 9 spectra of
4 different O stars (used to help locate telluric water vapor bands in
the red spectra; Wade \& Horne 1988\markcite{wade88}).  Finally,
George Jacoby kindly took two spectra of 4U 1543-47, the spectrum of
an A0V comparison star, and the spectrum of a flux standard using the
Blanco 4 meter telescope in photometric conditions on the night of
1997 June 29.  He used the spectrograph, camera, and detector
mentioned above, and the \#181 grating ($316\ell$/mm, first order).
These spectra cover 4700 to 10,028~\AA\ in 2.0~\AA\ pixels, giving a
spectral resolution of about 6~\AA\ FWHM.

IRAF tasks were used to remove the electronic bias and to perform the
flat-field corrections of the 2-D spectra.  Then optimal extractions
(Horne 1986\markcite{horne86}) of the spectra were done with the
programs in the CTIOSLIT package within IRAF.  The profile of 4U
1543-47 was well resolved from the profiles of its near neighbors (see
below), and the quality of our spectra is generally quite high.  The
signal-to-noise (S/N) ratios of our blue 4U 1543-47 spectra are 30 to
40 per pixel near 5000~\AA, while the S/N ratios of the red spectra
are 40 to 60 per pixel near 7000~\AA; however a few of the most weakly
exposed spectra from July 4 have S/N ratios as low as 20 per pixel.
The S/N ratios for the bright comparison A-type stars are in excess of
150 per pixel.

\section{Period Determination}\label{per}

\subsection{Spectroscopic Period}\label{rv}

\begin{figure*}[p]
\epsscale{1.6}
\plotone{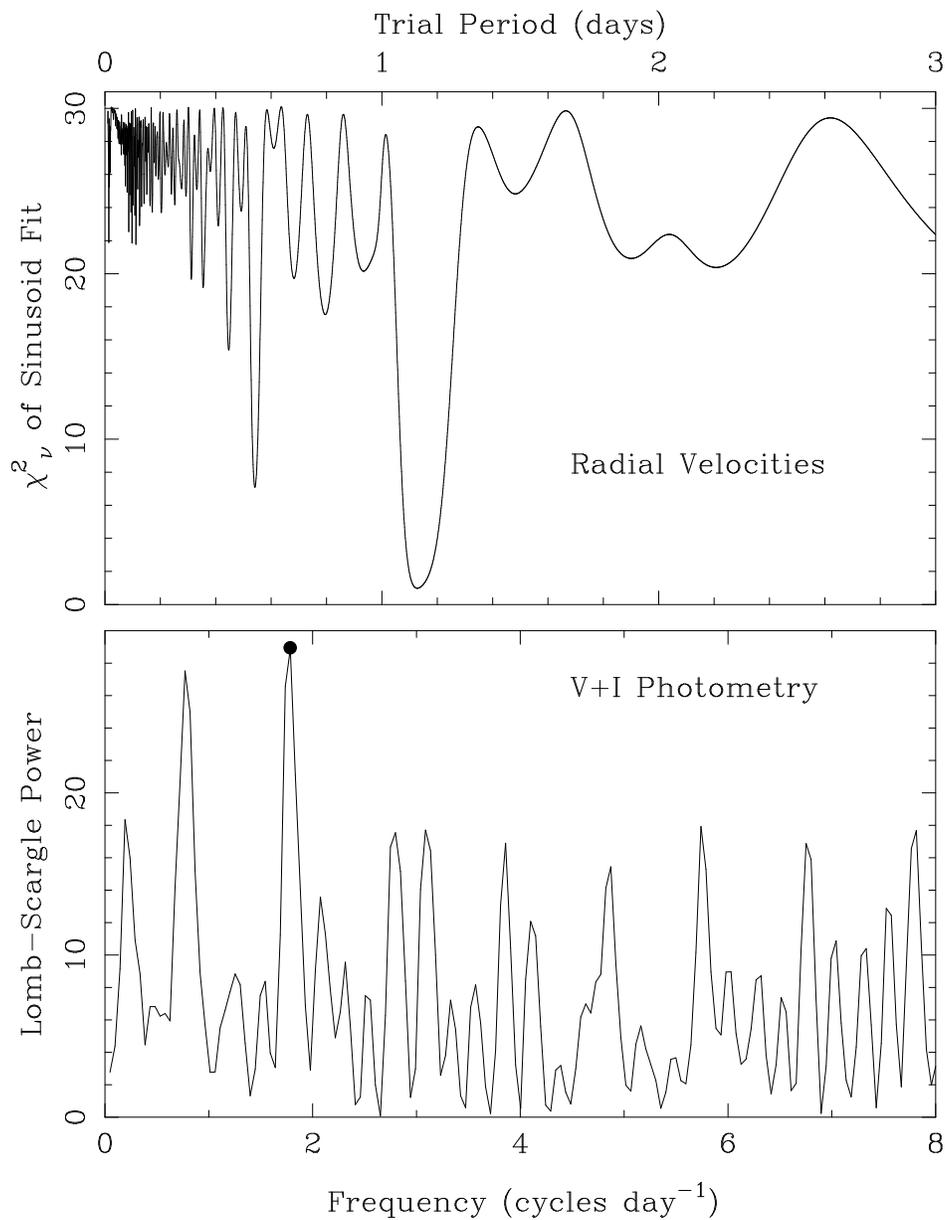}  
\caption{Top: The value of $\chi^2_{\nu}$ of a sinusoid fit to the 43
radial velocities as a function of the trial period.  The best fitting
sinusoid has a period of $P=1.123$ days.  Bottom: A Lomb-Scargle
periodogram for the combined $V+I$ light curve.  The highest peak
(marked with a filled circle) has a frequency of 1.784 cycles
day$^{-1}$, corresponding to a period of $P=0.5605$ days $=P_{\rm
spect}/2$.  The ``false-alarm probability'' for this peak is
$1.3\times 10^{-10}$.}
\label{fig0}
\end{figure*}

The radial velocities were determined from the spectra by
cross-correlating an ``object'' spectrum with a ``template'' spectrum
as described in Tonry \& Davis (1979\markcite{td79}).  The
calculations were done using FXCOR within IRAF.  For the blue spectra,
the cross-correlation (CC) functions were computed simultaneously over
the wavelength intervals 3750-4455, 4525-5750, and 6310-6715~\AA,
which includes all of the Balmer lines up to H12, and excludes some
strong interstellar lines such as the NaI D lines.  The CC functions
for the red spectra were computed over the wavelength intervals
6310-6840, 7000-7100, 7350-7550, 7700-8050, and 8370-8700~\AA.  We
used observations of the O stars at several different airmasses to
locate the strongest telluric water vapor lines in the red spectra
(Wade \& Horne 1988\markcite{wade88}), and the CC region was carefully
picked to avoid the telluric bands.  The strongest A-star lines in
this CC region are H$\alpha$ and lines 13 to 17 of the Paschen series.

The radial velocities of 4U 1543-47 were measured using the A0V star
HD 130095 as the template.  This comparison star was observed by us
six times over the four nights and was also observed by Jacoby the
night of June 29.  We adopted the first observation from July 1 as the
template for the blue spectra and the first observation from July 2 as
the template for the red spectra.  For every object spectrum the CC
was quite strong with a peak height ranging from 0.55 to 0.80.  The
velocity corresponding to the maximum of the CC functions was
determined by making a parabolic fit to the seven pixels centered on
the peak.  Typical values for the ``r'' parameter (Tonry \& Davis
1979\markcite{td79}) ranged from 15 to 25.

\begin{figure*}[t]
\plotfiddle{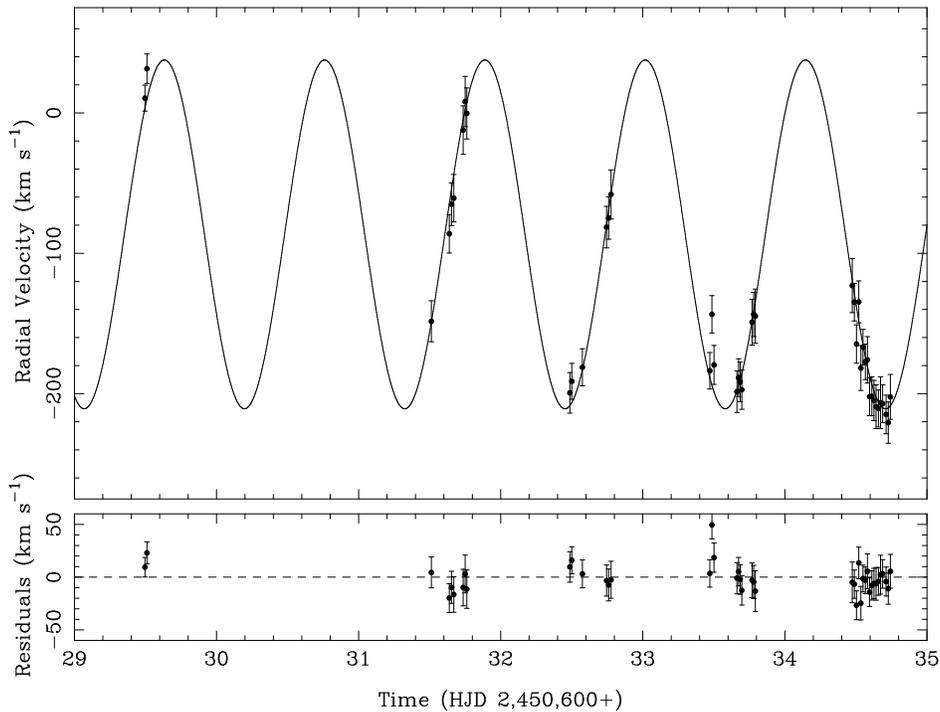}{3.5in}{-90.0}{50}{50}{-200}{300}
\caption{Top: The radial velocities of 4U 1543-47 and the best-fitting
sinusoid (solid line).  The errors on the velocities have been scaled
so that reduced chi-square of the sinusoid fit is 1.0.  Bottom: The
residuals in the sense of the data {\em minus} the model fit.}
\label{fig1}
\end{figure*}

The spectroscopic period was found by fitting a three-parameter
sinusoid to the resulting 43 velocities over a range of trial periods.
We then looked for the trial period $P$ which resulted in the lowest
value of $\chi^2_{\nu}$.  The three free parameters are the epoch of
maximum velocity $T_0$, the semi-amplitude $K_2$, and the systemic
velocity $\gamma$.  Figure \ref{fig0} shows how $\chi^2_{\nu}$ varies
as a function of the trial period.  The best sinusoid fit to the
radial velocities has a period of $P=1.123$ days, where
$\chi^2_{\nu}=0.448$.  The one day alias period ($P=0.529$ days) is
clearly ruled out, since $\chi^2_{\nu}=5.75$ at that period.  The
errors on the radial velocities were scaled to give $\chi^2_{\nu}=1$,
and the uncertainties on the fitted parameters were computed using the
scaled errors.  The period of $P=1.123\pm 0.008$ days and the
semi-amplitude of $K_2=124\pm 4$~km~s$^{-1}$ combine to give a mass
function of $f(M)=0.22 \pm 0.02\,M_{\sun}$.  Table~\ref{tab1} gives
the orbital parameters and Figure \ref{fig1} shows the velocities and
the best-fitting sinusoid.

\begin{deluxetable}{cc}
\tablewidth{0pt}
\tablecaption{Parameters for 4U 1543-47}
\tablehead{
\colhead{parameter}             &
\colhead{value}}
\startdata
Spectroscopic period $P$ (days)   &   $1.123 \pm 0.008$  \nl
$T_0$ (HJD 2,450,000$+$)          &   $629.6329\pm 0.0083$  \nl
$K_2$ velocity (km~s$^{-1}$)        &   $124\pm 4$          \nl
$\gamma$ velocity (km~s$^{-1}$)        &   $-87\pm 3$          \nl
$f(M)$ ($M_{\sun}$)               &   $0.22\pm 0.02$   \nl
\enddata
\label{tab1}
\end{deluxetable}

One must be cautious when dealing with periods near one day.  Flexure
in the spectrograph could in principle produce a spurious signal if
one observes a source at the same hour angles each night.  However, an
examination of the residuals (the data {\em minus} the model fit)
shown in Figure \ref{fig1} shows that there are no systematic trends
evident.  Furthermore, our period search excludes values near 1.0
days, as shown in the upper panel of Figure \ref{fig0}.  Based on our
past experience with this spectrograph and the reliable results we
have obtained (McClintock \& Remillard 1990\markcite{mr90}; Remillard,
McClintock, \& Bailyn 1992\markcite{rmb92}; Bailyn et al.\
1995b\markcite{bomr95b}; Remillard et al.\ 1996\markcite{rem96}; Orosz
et al.\ 1996\markcite{obmr96}), we are confident that $P=1.123\pm
0.008$ days is the true orbital period of 4U 1543-47.

\subsection{Photometric Period}

It is relatively difficult to obtain precise photometry of 4U 1543-47
since there is a star 7\farcs 25 to the southeast that is $\approx
4.5$ mag brighter in $V$ than the counterpart.  In addition, there is
a faint red star 2\farcs 5 to the northeast of the counterpart which
is $\approx 4$ magnitudes fainter in $V$.  Simple aperture photometry
would be inadequate, so we used the programs DAOPHOT IIe, ALLSTAR and
DAOMASTER (Stetson 1987\markcite{st87}; Stetson, Davis, \& Crabtree
1991\markcite{sdc91}; Stetson 1992a\markcite{st92a},b) to compute the
photometric time series of 4U 1543-47 and the $\approx 1000$ other
stars in the field.  DAOPHOT was used to construct an empirical point
spread function (PSF) for each frame from fits to relatively isolated
bright stars.  Then the PSFs were used to deblend close stars, thereby
making it possible to do accurate photometry in the crowded field.
The exposure times were adjusted so that only a few pixels of the
bright neighbor star were saturated, while leaving the image of 4U
1543-47 relatively well-exposed.  DAOPHOT and ALLSTAR were then used
to remove the wings of the bright star's profile.  The resultant S/N
ratios in the images of 4U 1543-47 were still adequate for precise
photometry. The DAOMASTER program assembles the photometric time
series from the individual photometry files by effectively using as
comparison stars all of the stable stars that are common to all of the
images.  In this way the derived light curves are not sensitive to the
fluctuations of a few comparison stars.

Both the $V$ and $I$ light curves of 4U 1543-47 show statistically
significant variations.  For example, the standard deviation (from a
constant mean value) of the $V$-band light curve of 4U 1543-47 is
0.038, compared to $0.005$ to $0.01$ for the $\approx 30$ field
comparison stars with brightnesses similar to that of 4U 1543-47.  The
amplitude of the $I$ light curve is about 0.015 magnitudes less than
the amplitude of the $V$ light curve.  We combined the $V$ and $I$
photometry after subtracting the mean from each set and searched for
periodicities by computing the Lomb-Scargle periodogram (Figure
\ref{fig0}) using an implementation due to Press et al.\
(1992)\markcite{press}.  The highest peak in the periodogram for the
combined $V+I$ band data has a frequency of 1.784 cycles per day,
corresponding to a period of 0.5605 days.  This is one-half of the
spectroscopic period (to within the error of $P_{\rm spect}$), as
expected for an ellipsoidal light curve with two maxima and two minima
per orbital cycle.  This main peak is about 60\% taller than all of
the other peaks (except for the one day alias peak at 0.784 cycles per
day) and has a ``false-alarm probability'' of $1.3\times 10^{-10}$,
where a small value of this probability indicates that significant
periodic signal is present (Press et al.\ 1992)\markcite{press}.

We expect to see an ellipsoidal light curve with a period equal to the
spectroscopic period because the secondary is Roche lobe-filling
(i.e.\ there are X-ray outbursts caused by accretion onto the compact
object) and is the dominate source of optical light in the system
(Section \ref{specclass}).  In several other black hole binaries the
{\em quiescent} photometric period (of the ellipsoidal modulation) has
been shown to be the same as the spectroscopic period derived from the
secondary star's absorption line radial velocity curve: A0620-00
(McClintock \& Remillard 1986\markcite{mr86}; Orosz et al.\
1994\markcite{o94}), GRS 1124-683/XN Mus91 (Orosz et al.\
1996\markcite{obmr96}; Casares et al.\ 1997\markcite{cas97}),
H1705-250/XN Oph77 (Remillard et al.\ 1996\markcite{rem96}; Filippenko
et al.\ 1997\markcite{fil97}), GRO J1655-40 (Orosz \& Bailyn
1997\markcite{ob97}; van der Hooft et al.\ 1997\markcite{vdh}), and
GRO J0422+32 (Orosz 1996\markcite{o96}; Filippenko, Matheson, \& Ho
1995\markcite{fil95}).  Based on these considerations and given the
fact that the highest peak in the periodogram has a period of $P_{\rm
spect}/2$, we are justified in asserting that $P_{\rm photo}=P_{\rm
spect}$ for 4U 1543-47.

The traditional definition of the photometric phase is $T_0{\rm
(photo)} = T_0{\rm (spect)}+0.75P$ (i.e.\ $T_0{\rm (photo)}$
corresponds to the time of the closest approach of the A-star).  We
estimated $T_0{\rm (photo)}$ in two different ways.  A sinusoid with a
period of 0.5605 days fit to the $V$ band data gives $T_0{\rm
(photo)}={\rm HJD}\, 2,450,629.380\pm 0.013$ (although in this case
there is a half cycle ambiguity).  A fit using an ellipsoidal model
(see Section \ref{incl} below) gives $T_0{\rm (photo)}={\rm HJD}\,
2,450,629.362\pm 0.012$ (there is no half cycle ambiguity in this
case).  We adopt a value that is the average of the two: $T_0{\rm
(photo)}={\rm HJD}\, 2,450,629.37\pm 0.01$.  The spectroscopic phase
of $T_0{\rm (photo)}$ is $0.733\pm 0.013$, consistent with the
expected value of 0.75.  This agreement between the photometric and
spectroscopic phases has an a priori probability of $\approx 0.1$,
which further increases the significance of the peak at 0.5605 days.

\begin{figure*}[p]
\epsscale{1.6} 
\plotone{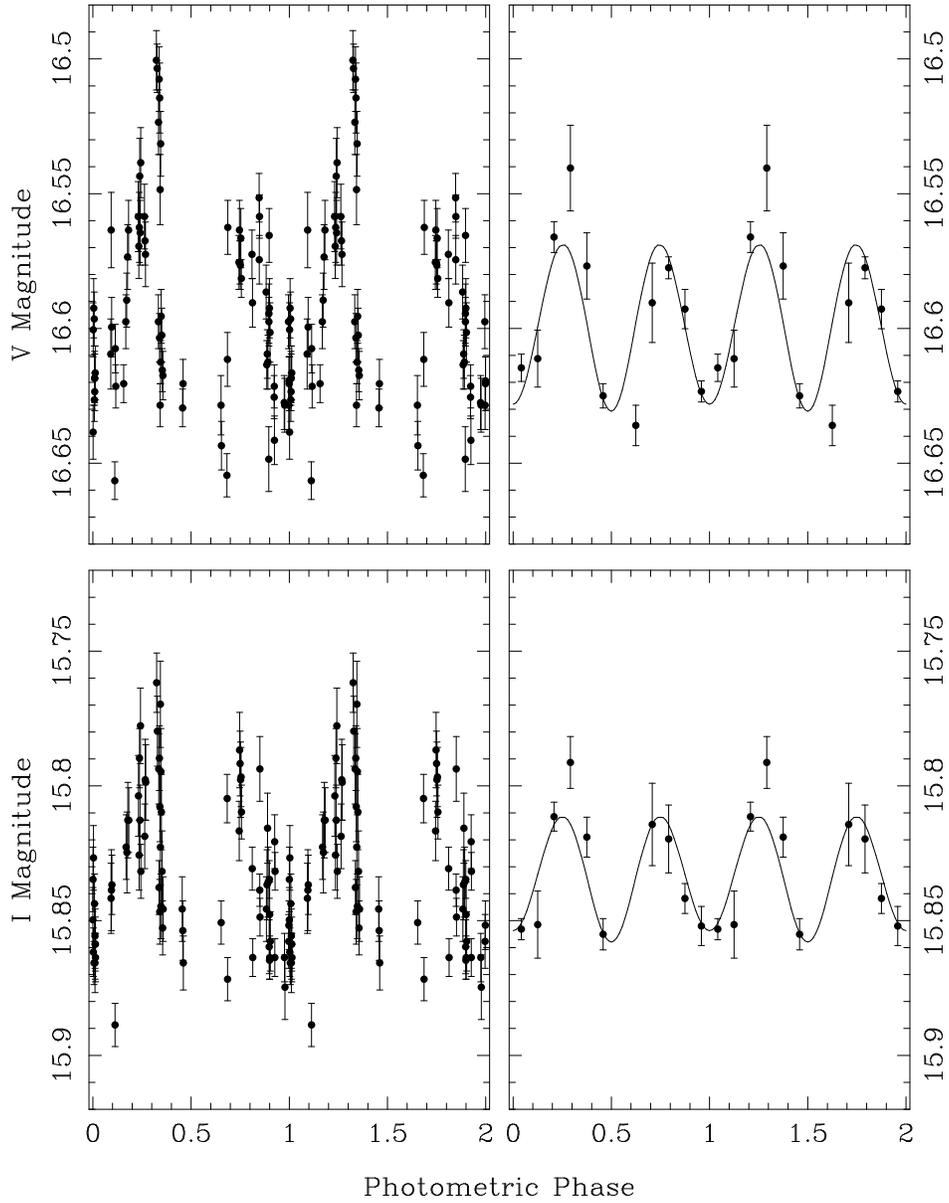} 
\caption{The left panels show the $V$ and $I$ light curves of 4U
1543-47 folded on the photometric phase, where $T_0{\rm (photo)} =
T_0{\rm (spect)}+0.75P$.  The right panels show the folded light
curves binned into 12 phase bins.  Note that a few of the bins are
empty.  The curves shown are an ellipsoidal model
with an inclination of $31\arcdeg$ and a mass ratio of 2.}
\label{fig2}
\end{figure*}

The $V$ and $I$ light curves of 4U 1543-47 (corrected to the standard
magnitude scales) are shown in Figure \ref{fig2} folded on the
photometric phase.  The right panels in Figure \ref{fig2} show the
light curves binned into 12 phase bins (where a few of the bins are
empty).  Although the light curves are somewhat noisy, they resemble
an ellipsoidal light curve because they have maxima at phases
consistent with 0.25 and 0.75 and minima at phases consistent with 0.0
and 0.5.  The amplitudes of the light curves are small, roughly 0.08
magnitudes in $V$ and 0.07 magnitudes in $I$.  The ellipsoidal models
shown with the binned light curves are discussed below in Section
\ref{incl}.  Note that the existence of ellipsoidal variations implies
that the A star is the mass donor and not the outer star of a triple
system.

\section{Spectral Classification and Reddening}\label{specclass}

\begin{figure*}[t]
\plotfiddle{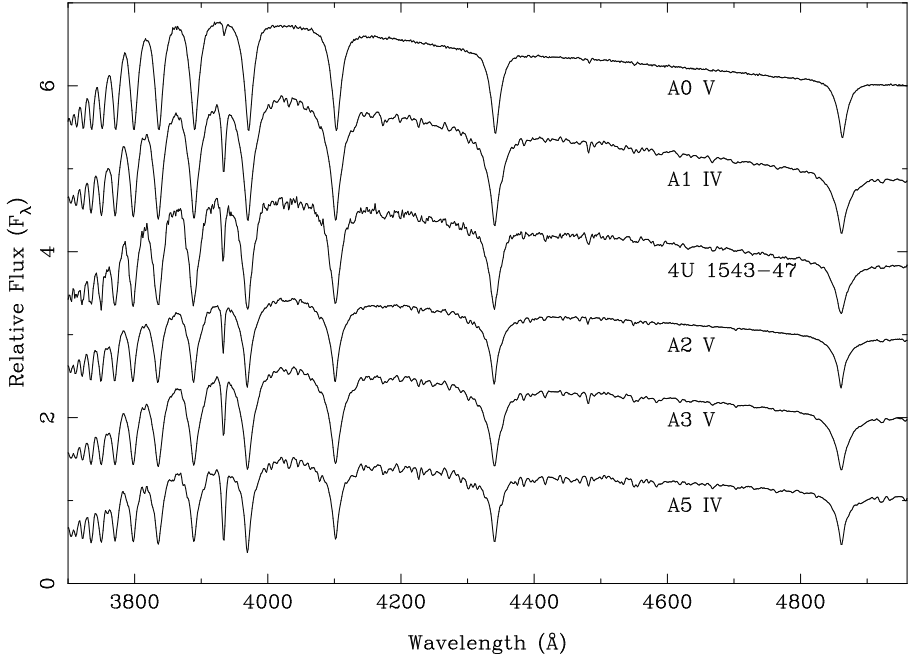}{3.0in}{-90.0}{50}{50}{-220}{300}
\caption{The dereddened blue restframe spectrum of 4U 1543-47 is shown
along with the spectra of several A stars.  All of the flux-calibrated
spectra have been normalized to unity at 5000~\AA\ and offsets have
been applied for clarity.  }
\label{fig3}
\end{figure*}

Figure \ref{fig3} shows the blue ``restframe'' spectrum of 4U 1543-47
and the spectra of several A-type comparison stars.  (The restframe
spectrum is a sum of the individual spectra, which have been corrected
to zero velocity).  The spectrum of 4U 1543-47 was dereddened using
$E(B-V)=0.5$ (see below).  The \ion{Ca}{2} K line at 3933~\AA\ is a
strong temperature indicator in A stars, and its equivalent width of
2.1~\AA\ gives a spectral type of A2-A3 (Jaschek \& Jaschek
1987\markcite{jasjas}).  Some of the K line could be due to
interstellar absorption, in which case the spectral type would be
slightly earlier.  In principle, the widths of the Balmer lines can be
used as a luminosity indicator in A stars (Jaschek \& Jaschek
1987\markcite{jasjas}), but the difference in widths between dwarfs
and subgiants is too small for us to measure.  We will adopt a
spectral type of A2V, the same type determined by Chevalier
(1989)\markcite{chev89}.

Our broad spectral coverage affords us a good opportunity to make an
independent estimate of the reddening.  The precision of our absolute
flux calibration was compromised by the use of a narrow slit; however,
the relative spectrophotometry should be much better since we rotated
the spectrograph slit to match the parallactic angle, thereby
minimizing the effects of differential atmospheric refraction
(Filippenko 1982\markcite{fil82}).  Indeed, the flux values of the
restframe spectra from June 29 and July 1 agreed to better than 1\%
between 4700 and 6560~\AA\ after each spectrum was normalized to unity
at 5000\AA\ (there was about a 5\% difference between the two spectra
between the H$\alpha$ line and the telluric A band).  We averaged the
two normalized restframe spectra to create a continuous spectrum from
3600 to 9100~\AA.

\begin{figure*}[t]
\plotfiddle{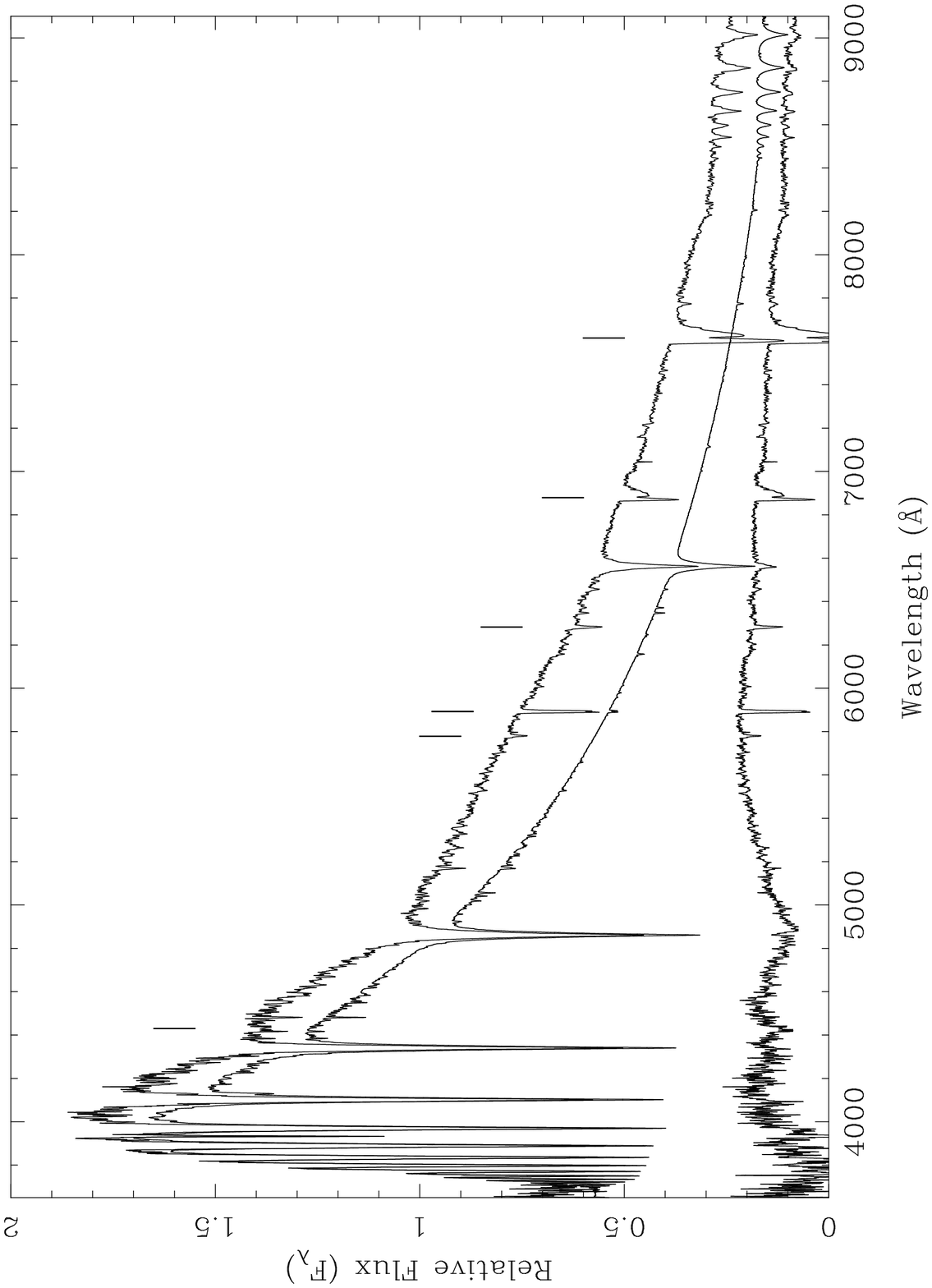}{3.0in}{-90.0}{50}{50}{-220}{300}
\caption{The normalized dereddened restframe spectrum of 4U 1543-47
made by averaging the spectra from 1997 June 29 and July 1 is shown on
the top.  The spectrum in the middle is a Kurucz model spectrum with
$T_{\rm eff}=9500$~K and $\log g=4.0$ scaled by 0.9.  The difference
spectrum is shown at the bottom.  Strong interstellar and telluric
lines are indicated with the vertical bars.}
\label{fig4}
\end{figure*}

We used synthetic spectra generated from Kurucz models with solar
metallicity to decompose the spectrum of 4U 1543-47 into its stellar
and disk components.  The observed A-star comparisons made unsuitable
templates because of the broad temperature range and the large line
broadening due to rapid rotation.  The decomposition technique we used
has been discussed elsewhere (Marsh, Robinson, \& Wood,
1994\markcite{mrw94}; Casares \& Charles 1994\markcite{cas94};
OB97\markcite{ob97}).  Basically a normalized template spectrum is
scaled by a factor $w$, subtracted from the normalized restframe
spectrum, and the scatter in the difference spectrum is measured by
computing the rms (per 1.2~\AA\ bin) from a polynomial fit (with
interstellar and telluric lines masked out of the fit).  The process
is repeated until the ``smoothest'' difference spectrum is found
(corresponding to the lowest overall rms from the polynomial fit).
The smoothest difference spectrum is taken to be the spectrum of the
disk.  The relative contributions of the star and disk to the total
system flux can be measured by looking at the relative levels of the
restframe spectrum and the scaled template spectrum.  In our case we
first dereddened the restframe spectrum by various amounts and used
synthetic template spectra characterized by several different
effective temperatures ($T_{\rm eff}$) and surface gravities ($\log
g$).  We then looked for the combination of $E(B-V)$, $T_{\rm eff}$,
and $\log g$ that gave the smoothest overall difference spectrum.  The
best combinations were for $8500 < T_{\rm eff} < 9500$~K, $3.5 < \log
g < 4.0$, and $0.45 < E(B-V) < 0.55$.  Figure \ref{fig4} shows an
example of a decomposition where the restframe spectrum was dereddened
using $E(B-V)=0.5$ and where the synthetic spectrum has $T_{\rm
eff}=9500$ and $\log g=4.0$.  The temperature range of $8500 < T_{\rm
eff} < 9500$~K corresponds roughly to a spectral type range of A1 to
A3 (Gray 1992\markcite{gray}).

For comparison, our derived range of values for the color excess
($0.45 < E(B-V) < 0.55$) is slightly lower than the range derived from
X-ray observations.  The hydrogen column density $N_H$ estimated from
X-ray spectra ranges from $3.1\times 10^{21}$~cm$^{-2}$ (Greiner et
al.\ 1994\markcite{grien94}) to $(4.26\pm 0.15)\times
10^{21}$~cm$^{-2}$ (van der Woerd et al.\ 1989\markcite{vdW89}).  The
implied values of the color excess $E(B-V)$ range from 0.56 to 0.77
assuming $A_V=N_H/1.79\times 10^{21}$ (Predehl \& Schmitt
1995\markcite{ps95}) and $E(B-V)=A_V/3.1$.  Note that there may be a
significant contribution to $N_H$ due to X-ray absorption in 4U
1543-47 itself, in which case the derived values of $N_H$ and $E(B-V)$
would be inflated.

The disk spectrum shown in Figure \ref{fig4} is approximately flat
across the entire range, although the relatively poor S/N at the blue
end makes it difficult to assess the precise spectral shape there.
With the exception of H$\alpha$, the disk spectrum is free of hydrogen
lines in absorption, and is also devoid of emission lines.  The
accretion disk contributes roughly 10\% of the flux in $B$, 21\% in
$V$, 32\% in $R$, and 39\% in $I$ (the uncertainties in these values
are $\lesssim 5\%$).  Although the disk spectrum will be used as input
to the ellipsoidal models discussed in Section \ref{incl}, we will not
discuss the disk spectrum further.  A thorough discussion of the
relative faintness of the disk, the lack of emission lines, and the
overall spectral shape is beyond the scope of this paper.

\section{Inclination Constraints}\label{incl}

The amplitude of an ellipsoidal light curve depends strongly on the
inclination of the binary orbit (Avni 1978\markcite{av78}).  In
practice, however, one must carefully assess the strength of extra
sources of light, such as light from the disk.  The small amplitudes
of the light curves for 4U 1543-47 imply a small inclination (the disk
is relatively faint and does not substantially reduce the amplitude of
the observed light curves).  We modeled the light curves using the
modified version of the Avni (1978\markcite{av78}) code described in
OB97\markcite{ob97} (also see Orosz \& Bailyn 1997b\markcite{ob97b}).
We assume the secondary star is in synchronous rotation and completely
fills its Roche lobe.  The effective temperature of the secondary was
taken to be 9000K, and the limb darkening coefficients for a $\log
g=4$ atmosphere were taken from Wade \& Rucinski
(1985)\markcite{wr85}.  The A-star has a radiative envelope, so the
gravity-darkening exponent was set to 0.25 (see OB97\markcite{ob97}).
For 4U 1543-47, which is a non-eclipsing system, we used the
empirically determined disk spectrum (Figure \ref{fig4}) as input to
the modified Avni code.  The continuum fit of the disk spectrum
$D(\lambda)$ was divided by the continuum fit of the scaled stellar
spectrum $S(\lambda)$, producing the ``ratio spectrum'' $R(\lambda)$.
The code then determines the model disk spectrum for all phases by
multiplying $R(\lambda)$ by the model star's spectrum at phase 0.0.
The model disk spectrum and the model star spectrum at each phase are
summed and integrated with the filter response functions to produce
the model flux in the five filters.  This method of computing fluxes
at several different wavelengths and then integrating with the filter
response functions to get the final $UBVRI$ fluxes is much more
accurate than simply computing the flux at the ``effective''
wavelength of each filter because the effective wavelength of the
filter bandpasses depends on the shape of the input spectrum.

X-ray heating of the secondary star can alter the shape of the
observed light curve when $L_x\gtrsim L_{\rm opt}$ (e.g.\ Avni
1978\markcite{av78}).  4U 1543-47 was observed in the quiescent state
1996 February with the SIS detector on the {\em Advanced Spacecraft
for Cosmology Astrophysics} (ASCA).  We reanalyzed the publicly
available data and found that the source was not detected after a 20.4
ksec observation, indicating an X-ray flux limit (0.5 - 10 keV) of
$F_x\lesssim 2\times 10^{-13}$ ergs s$^{-1}$ cm$^{-2}$ ($\approx
5\sigma$), assuming a power law spectrum with an index of 2 and a
hydrogen column density of $N_H=3.5\times 10^{21}$ cm$^{-2}$.  The
corresponding X-ray luminosity limit is $L_x\lesssim 3.83\times
10^{32}(d/4\, {\rm kpc})^2$ ergs s$^{-1}$, a typical value for a
quiescent X-ray transient (Tanaka \& Shibazaki 1996\markcite{tanaka}).
We therefore neglect X-ray heating since the quiescent X-ray
luminosity of the system is a very small fraction of the bolometric
luminosity of the A-star secondary: $L_{\rm bol}\gtrsim 95L_x$ for a
distance of 9 kpc (Section \ref{dist}).

\begin{figure*}[t]
\plotfiddle{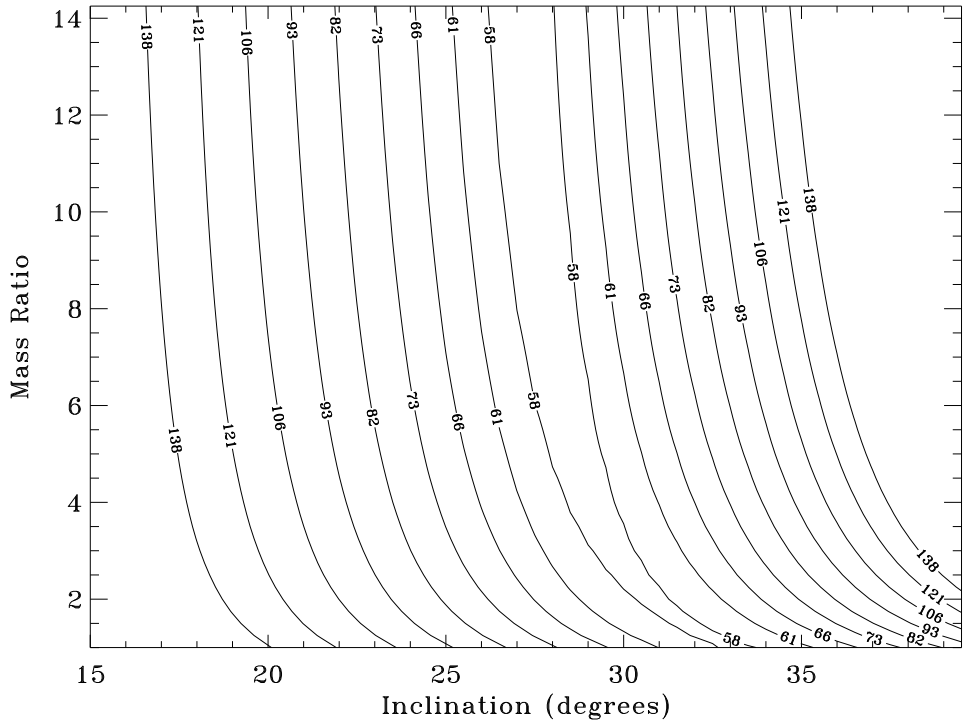}{3.5in}{90.0}{60}{60}{250}{-25}
\caption{$\chi^2$ contours for simultaneous ellipsoidal model fits to the $V$
and $I$ binned light curves.  There are 11 points in the $V$ light
curve and 10 points in the $I$ light curve.  The minimum $\chi^2$
value is 57.32.  The contours levels are $\chi^2_{\rm min}+1$,
$\chi^2_{\rm min}+4$, $\chi^2_{\rm min}+9$, etc.  The formal $3\sigma$
inclination range is $24\le i \le 36\arcdeg$ while the mass ratio is
not well-constrained.  A fit to the unbinned light curves produces a
similar contour plot.}
\label{fig5}
\end{figure*}

Once the disk light is accounted for, the two remaining free
parameters for the ellipsoidal model are the inclination $i$ and the
mass ratio $Q\equiv M_1/M_2$, where $M_1$ is the mass of the compact
primary and $M_2$ is the mass of the A-star secondary.  Note that the
individual component masses themselves are not important---the
relative shapes of the Roche lobes depend only on the mass {\em
ratio}.  The binned $V$ and $I$ light curves were fit simultaneously,
producing the $\chi^2$ contours in the $Q\,{\em vs.}\,i$ plane shown
in Figure \ref{fig5}.  The inclination is tightly constrained between
$24\arcdeg$ and $36\arcdeg$ (the formal $3\sigma$ limits for $Q>1$
corresponding to the $\chi^2=\chi^2_{\rm min}+9=66$ contours), while
the mass ratio is basically unconstrained (i.e.\ the $\chi^2$ contours
are nearly parallel to the $y$-axis).  We argue that the A-star is on
the main sequence (see Section \ref{disc} below), which implies mass
ratios near $\approx 2$.  For mass ratios near 2, the $\chi^2=66$
contours enclose an inclination range of $27\le i\le 36\arcdeg$.
However, since we cannot immediately rule out the larger mass ratios,
we adopt the wider range of inclinations: $24\le i\le 36\arcdeg$.

The model light curves for $i=31\arcdeg$ and $Q=2$ are shown as the
solid lines in Figure \ref{fig2}.  Except for some excess light near
phase 0.25, the $i=31\arcdeg$ models fit the light curves reasonably
well.  Excess light has been seen in the light curves of Nova Muscae
1991 (Orosz et al.\ 1996\markcite{obmr96}) and GRO J0422+32 (Casares
et al.\ 1995\markcite{cas95}; Orosz 1996\markcite{o96}).  Orosz et
al.\ (1996\markcite{obmr96}) argued that the asymmetry in the light
curves of Nova Muscae 1991 was due to extra phase-modulated light from
the accretion disk and that a model fit to the smaller maximum and the
two minima should be more trustworthy.  If the asymmetry of the 4U
1543-47 light curves is also due to phase-modulated light from the
disk, then the model fits to the 4U 1543-47 light curves shown in
Figure \ref{fig2} should be fairly reliable.  In any event, the
presence of the excess light near phase 0.25 and the large
$\chi^2$/DOF ($\approx 2.9$ at the minimum) does indicate unaccounted
for systematic effects, so the inclination limits quoted above should
be treated with caution.  Light curves with better phase coverage and
higher statistical precision will be needed before definitive
inclination limits can be established.

We can derive a lower limit on the inclination by fitting a model that
has no disk contamination.  A model with $i=20\arcdeg$ and $Q=10$ has
an amplitude between phases 0.75 and 1.0 which is $\approx 15\%$
smaller than the amplitudes of the observed light curves in the same
phase interval.  Since the addition of disk light will act to reduce
the amplitude of the observed light curve, we can adopt $i\ge
20\arcdeg$ as a reasonable lower limit on the inclination since the
inclination of the model will need to increase in order to match the
observations.  Similarly, a model with $i=40\arcdeg$, $Q=2$, and a
21\% disk fraction in $V$ (i.e.\ the observed value) has an amplitude
between phases 0.25 and 0.50 $\approx 15\%$ larger than the amplitudes
of the observed light curves in the same phase interval.  The
difference in amplitudes is $\approx 25\%$ between phases 0.75 and
1.0.  We will adopt $i\le 40\arcdeg$ as an upper limit on the
inclination.  These extreme inclination limits correspond roughly to
the $\chi^2=\chi^2_{\rm min}+49=106$ contours shown in Figure
\ref{fig5}.

\begin{figure*}[p]
\plotfiddle{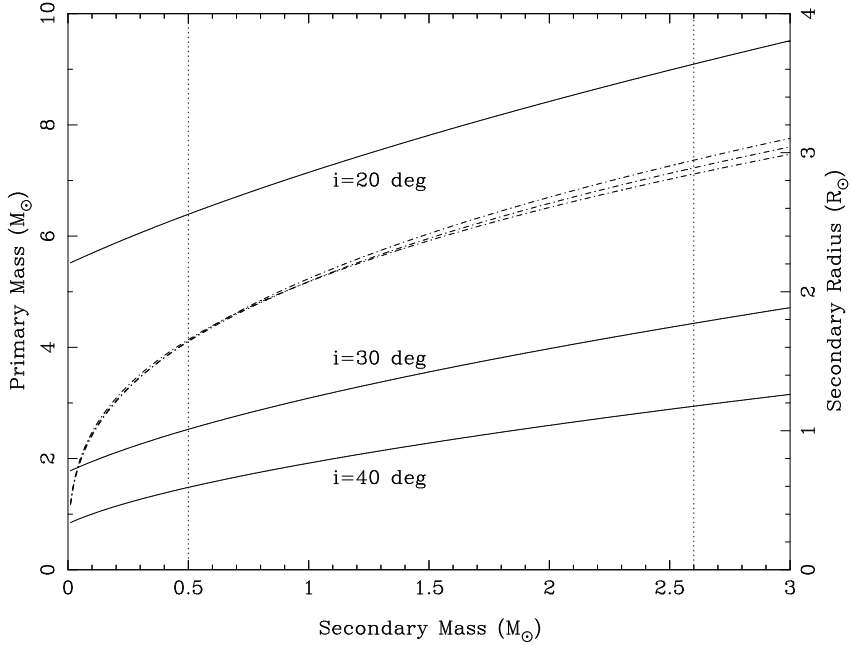}{2.3in}{-90.0}{45}{45}{-220}{260}
\caption{The mass of the primary $M_1$ (solid lines, left $y$-axis
scale) and the radius of the secondary $R_2$ (dash-dotted lines, right
$y$-axis scale) as a function of the secondary star mass $M_2$ for a
mass function of $f(M)=0.22\,M_{\sun}$ and $i=20$, 30, and $40\arcdeg$.
The vertical dotted lines indicate the extreme range of secondary star
masses $0.5\le M_2\le 2.6\,M_{\sun}$.}
\label{fig6}
\end{figure*}
\begin{figure*}[p]
\plotfiddle{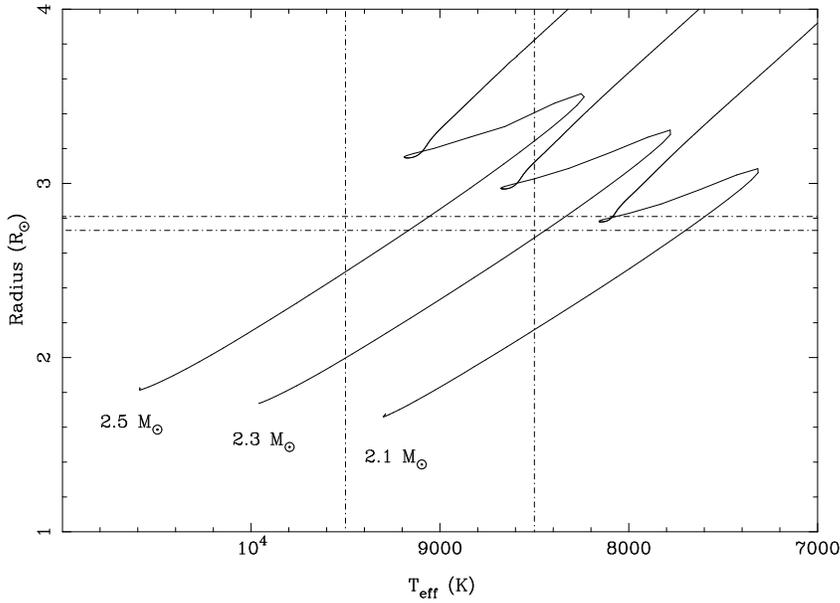}{2.3in}{-90.0}{45}{45}{-220}{260}
\caption{The radius of the model stars as a function of the effective
temperature. The stars begin H burning at the lower left and move to
larger radii and cooler temperatures as they age.  Core H burning
stops just beyond the tiny loops in the tracks.  The vertical lines
indicate the range of temperatures measured for the 4U 1543-47
secondary, and the horizontal lines correspond to the range of radii
of a star with $2.1\le M_2\le 2.5\,M_{\sun}$ (see Figure
\protect\ref{fig6}).}
\label{fig7}
\end{figure*}

\section{Discussion}

\subsection{Constraints on the Component Masses}\label{disc}

One can easily compute the mass of the compact primary $M_1$ when
given the mass function $f(M)$, the inclination $i$, and the mass of
the A-star secondary $M_2$:
\begin{equation}
f(M)\equiv {PK_2^3\over 2\pi G}={M_1^3\sin^3i\over (M_1+M_2)^2}.
\label{equ1}
\end{equation}
Figure \ref{fig6} shows plots of $M_1$ as a function of $M_2$ for
$i=20$, 30, and $40\arcdeg$ (solid lines, left $y$-axis scale).  
We must now consider the possible values of the secondary star mass
$M_2$ in order to derive useful limits on the mass of the compact primary.

It is clear from the spectra (Figure \ref{fig3}) and the models
(Section \ref{specclass}) that the secondary star is an early A-star
with $T_{\rm eff}\approx 9000$ K.  If the A-star is on the main
sequence, then we would expect its mass to be near $\approx
2.2\,M_{\sun}$ (Gray 1992\markcite{gray}).  On the other hand, the
secondary star could be a highly evolved star on the extended
horizontal branch (EHB) where it would have a He burning core and a
thin H envelope and a mass of $\approx 0.5\,M_{\sun}$.  Initially, one
might expect that the main sequence star would have a rather different
surface gravity than the EHB star, which would lead to rather
different Balmer line profiles (e.g.\ Jaschek \& Jaschek
1987\markcite{jasjas}).  However, the following simple exercise shows
that the surface gravity ($\log g$) of the secondary star varies by
only a small amount over a wide range of masses $M_2$.  The mean
density of the secondary star depends only on the orbital period to a
very good approximation (Pringle 1985\markcite{pring85}), i.e.\
$\rho_2\approx {\rm const}$.  This implies $R_2\propto M_2^{1/3}$
(Figure \ref{fig6}, dashed-dotted lines, right $y$-axis scale).  Since
$g\propto M_2/R_2^2$, $\log g\propto 1/3\log M_2$.  The difference in
the $\log g$ values between the main sequence star and the EHB star
would be less than 0.25, which is too small for us to measure using
our current spectra.

A strong indication that the secondary star probably is on the main
sequence comes from an examination of stellar models.  We computed the
evolutionary tracks of a $2.1\,M_{\sun}$ star, a $2.3\,M_{\sun}$ star,
and a $2.5\,M_{\sun}$ star using the Yale Stellar Evolution Code
(Guenther et al.\ 1992\markcite{gdkp92}) with updated opacities
(Iglesias \& Rogers 1996)\markcite{ir96}.  We assumed solar
metallicity, a hydrogen abundance of $71\%$, and a mixing length
coefficient of 1.7, all typical values for population I stars.  The
radius of these model stars as a function of the effective temperature
is shown in Figure \ref{fig7}.  The vertical lines indicate the likely
range of the temperature of the 4U 1543-47 secondary star, and the
horizontal lines indicate the range of radii of the secondary star
when $2.1\le M_2\le 2.5\,M_{\sun}$ (Figure \ref{fig6}).  The
$2.5\,M_{\sun}$ star appears on the zero-age main sequence (ZAMS)
hotter than the secondary star in 4U 1543-47.  However, the
$2.5\,M_{\sun}$ star will expand and cool {\em while on the main
sequence} until it attains a temperature of $\approx 9000$~K (the
observed temperature of the secondary star) and a radius of $\approx
2.8\,R_{\sun}$, which is precisely the radius the secondary star in 4U
1543-47 would have if $M_2=2.5\,M_{\sun}$ (Figure \ref{fig6}).  In
fact, stars with $2.3\lesssim M_2\lesssim 2.6\,M_{\sun}$ will pass
through the thin box in the center of Figure \ref{fig7} during their
main sequence lifetimes ($\approx 393$~Myr past the ZAMS in the case
of the $2.5\,M_{\sun}$ star).  One should note that the evolutionary
tracks shown in Figure \ref{fig7} are for {\em single} stars.  The
secondary star in 4U 1543-47 has been losing mass, and this is not
accounted for in the evolution models.  However, Kolb et al.\
(1997\markcite{kolb}) show that in the case of GRO J1655-40 (whose
secondary has a mass similar to that of 4U 1543-47 but is roughly
twice the age (OB97\markcite{ob97})), the proper binary evolution
models which account for the mass loss give similar results as the
single star evolutionary models.  Therefore the single star models
applied to 4U 1543-47 should also be a good approximation since it is
a younger system.

We conclude the secondary star is probably on the main sequence based
on the fact that the mean density of the secondary is consistent with
the mean density of a main sequence star with the same spectral type.
This conclusion does not depend on any assumed value of the
inclination since the computed radius of the secondary in 4U 1543-47
is a very weak function of the inclination.  We adopt a mass range of
$2.3\le M_2\le 2.6\,M_{\sun}$ for the A-star secondary.  Using these
main sequence values for the secondary star's mass, the best fits for
the $3\sigma$ inclination range ($24\le i\le 36\arcdeg$) and $3\sigma$
mass function range ($0.16\le f(M)\le 0.28\,M_{\sun}$) imply a primary
mass in the range $2.7\le M_1\le 7.5\,M_{\sun}$.

We can derive an extreme lower limit on the primary mass $M_1$ by
adopting the following extreme values of the parameters: $M_2 = 0.5
\,M_{\sun}$ and $i = 40 \arcdeg$.  In addition, we adopt $f(M) = 0.16
\,M_{\sun}$, which is the $3\sigma$ lower limit on the mass function,
and conclude: $M_1 \ge 1.2\, M_{\sun}$.  Thus the primary mass is at
least $1.2\,M_{\sun}$, and in fact is probably in excess of
$2.7\,M_{\sun}$ (i.e.\ the lower end of the range given above for main
sequence secondary).  The widely adopted maximum mass of a stable
neutron star is $M_{\rm max}\approx 3\,M_{\sun}$.  We therefore
conclude that this compact object is likely to be a black hole.

\subsection{Constraints on the Distance and Peak X-ray Luminosity}\label{dist} 

\begin{figure*}[t]
\plotfiddle{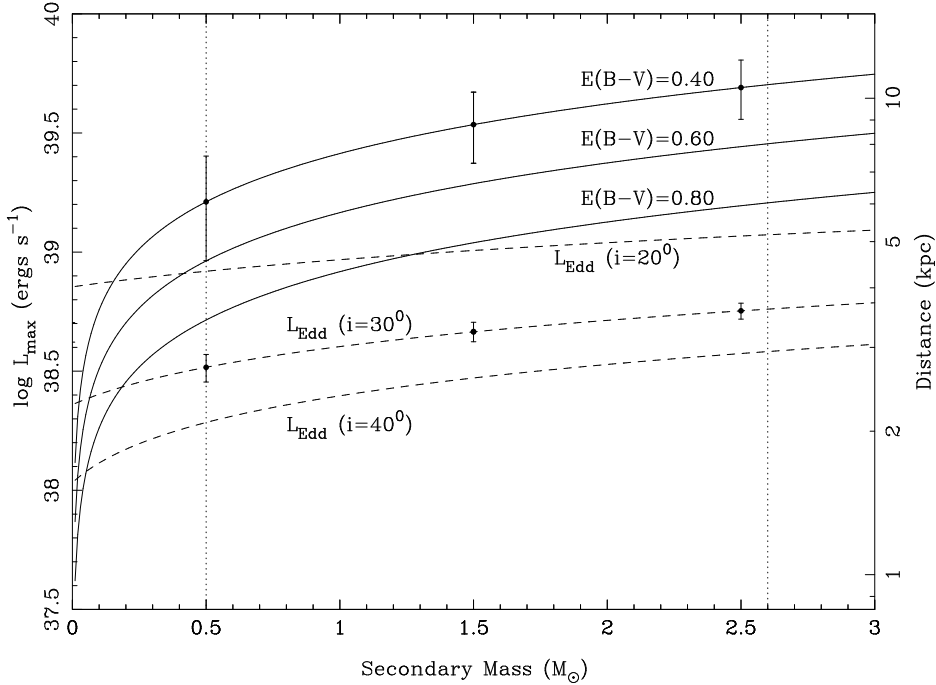}{3.5in}{-90.0}{50}{50}{-220}{300}
\caption{The solid lines show the peak X-ray luminosity during the
1983 outburst $L_{\rm max}$ as a function of the secondary star mass
$M_2$ (left $y$-axis scale) for three values of $E(B-V)$.  The right
$y$-axis scale shows the corresponding distances.  The dashed lines
show the Eddington luminosity as a function of $M_2$ for $i=20$, 30,
and $40\arcdeg$.  The points on the upper solid line show the
statistical error on $L_{\rm max}$ due to the errors on $T_{\rm eff}$,
$R_2$, and the disk fraction when $E(B-V)=0.40$.  The points on the
middle dashed line show the statistical error on $L_{\rm Edd}$ due to
the errors on the period $P$ and the mass function $f(M)$ when
$i=30\arcdeg$.}
\label{fig8}
\end{figure*}

We can easily compute the distance to the source once some basic
assumptions are made.  The distance $d$ depends mainly on the
secondary star mass $M_2$ and on the color excess $E(B-V)$; the
dependence of $d$ on the inclination $i$ is very weak.  The strong
dependence of $d$ on $M_2$ comes about because the orbital period
fixes the mean density of the Roche lobe-filling secondary star.  The
secondary star mass $M_2$ can then of course be combined with the
density to get the effective radius $R_2$.  For a main sequence
secondary we compute $R_2=2.84\pm 0.11\,R_{\sun}$ for $20\le i\le
40\arcdeg$ and $2.3\le M_2\le 2.6\,M_{\sun}$.  For the low mass end
($M_2=0.5\,M_{\sun}$) we compute $R_2=1.65\pm 0.07\,R_{\sun}$.  Once
the radius is known, we can take the well-determined temperature of
the secondary ($T_{\rm eff}=9000\pm 500$~K) and apply the
Stefan-Boltzmann relation to get the luminosity\footnote{Note that
standard tables relating spectral type to luminosity such as the one
given in Gray (1992) often give the absolute magnitude of the star
{\em when it is on the ZAMS}.  However, as shown in Figure
\protect\ref{fig7}, main sequence stars with similar temperatures can
have a large range of radii, resulting in a large range of
luminosities.  Thus our method of finding the intrinsic luminosity of
the secondary is more precise.}.  The distance modulus can be found
after light from the accretion disk ($21\pm 5\%$ of the total light in
$V$) is accounted for and the reddening correction is made.  There are
two measurements of $E(B-V)$ we will consider: $E(B-V)=0.50\pm 0.05$
derived by us and $E(B-V)=0.66\pm 0.10$ found from the range of $N_H$
determinations discussed above.  Table \ref{tab2} gives the 4 distance
determinations made from the 2 values of $R_2$ and the 2 values of
$E(B-V)$.  Our preferred value is $d=9.1\pm 1.1$ kpc (this assumes our
spectroscopically determined value of $E(B-V)$ and a main sequence
secondary), which is more than a factor of two larger than the
commonly assumed distance of 4 kpc (e.g.\ Chevalier
1989\markcite{chev89}).  The peak X-ray luminosity during the 1983
outburst is also given, where $L_{\rm max}=7.08\times 10^{38}
(d/4~{\rm kpc})^2$~ergs~s$^{-1}$ (conservatively estimated assuming a
thermal bremsstrahlung spectrum; Kitamoto et al.\
1984\markcite{kit84}).  The quoted errors on $d$ were computed
assuming the given errors on the input parameters are $1\sigma$
Gaussian errors.

\begin{deluxetable}{cccccc}
\tablewidth{0pt}
\tablecaption{Distance Determinations for 4U 1543-47\tablenotemark{a}}
\tablehead{
\colhead{{\begin{tabular}{c} Assumed Radius \\ ($R_{\sun}$)\end{tabular}}}    &
\colhead{$E(B-V)$}   &
\colhead{{\begin{tabular}{c} Secondary Luminosity \\ ($L_{\sun}$) \end{tabular}}}  &
\colhead{$M_{\rm bol}$}              &
\colhead{{\begin{tabular}{c} Distance \\ (kpc) \end{tabular}}}       &
\colhead{{\begin{tabular}{c} $\log L_{\rm max}$\tablenotemark{b} \\ (ergs~s$^{-1}$) \end{tabular}}}
}
\startdata
$2.84\pm 0.11$\tablenotemark{c} & $0.50\pm 0.05$ & $48\pm 11$ 
     & 0.51 & $9.1\pm 1.1$ & $39.56\pm 0.11$ \nl
$2.84\pm 0.11$ & $0.66\pm 0.10$ & $48\pm 11$ & 0.51 & $7.2\pm 1.0$ 
     & $39.36\pm 0.12$ \nl
$1.65\pm 0.07$\tablenotemark{d} & $0.50\pm 0.05$ & $16\pm 4$ 
     & 1.69 & $5.3\pm 0.6$ & $39.09\pm 0.11$ \nl
$1.65\pm 0.07$ & $0.66\pm 0.10$ & $16\pm 4$ & 1.69 & $4.2\pm 0.6$ 
     & $38.89\pm 0.12$ \nl
\enddata
\tablenotetext{a}{Other parameters are $T_{\rm eff}=9000\pm 500$~K, 
$\bar{V}=16.6$, and a disk fraction of $21\pm 5\%$ in $V$.}
\tablenotetext{b}{Peak X-ray luminosity during the 1983 outburst.}
\tablenotetext{c}{Corresponds to $2.3\le M_2\le 2.6\,M_{\sun}$.}
\tablenotetext{d}{Corresponds to $M_2=0.5\,M_{\sun}$.}
\label{tab2}
\end{deluxetable}

To better visualize how the distance $d$ (and hence $L_{\rm max}$)
depends on $M_2$ and $E(B-V)$, we show in Figure \ref{fig8} plots of
$\log L_{\rm max}\, {\em vs.}\, M_2$ for several values of $E(B-V)$.
We have assumed $T_{\rm eff}=9000$~K, $\bar{V}=16.6$, and a disk
fraction of 21\% in $V$.  We also show the Eddington luminosity
($L_{\rm Edd}= 1.3\times 10^{38}(M/M_{\sun})$~ergs~s$^{-1}$) as a
function of $M_2$ for $i=20$, 30, and $40\arcdeg$ (in contrast to
$L_{\rm max}$, $L_{\rm Edd}$ does depend on $i$ since $L_{\rm Edd}$
scales as the mass of the compact object).  To give some idea of the
uncertainties, we show the statistical error on $L_{\rm max}$ due to
the errors on $T_{\rm eff}$, $R_2$, and the disk fraction when
$E(B-V)=0.40$ for various values of $M_2$.  Likewise we show the
statistical errors on $L_{\rm Edd}$ due to the errors on the period
$P$ and the mass function $f(M)$ when $i=30\arcdeg$ for the same
values of $M_2$.  For most of the values of $M_2$, $i$, and $E(B-V)$,
the derived maximum X-ray luminosity during the 1983 outburst exceeds
the Eddington limit by an amount much larger than the errors.  For
example, for our preferred value of $d=9.1$ kpc, $L_{\rm max}=
3.6\times 10^{39}$ ergs s$^{-1}$.  In this case the mass of the black
hole is in the range $2.7\le M_1\le 7.5\, M_{\sun}$ (this is the mass
range which assumes a main sequence secondary), which implies
$3.1\lesssim L_{\rm max}/L_{\rm Edd} \lesssim 8.7$.  Unless the value
of $E(B-V)$ has been seriously {\em underestimated} (larger values of
$E(B-V)$ imply a closer system and a smaller value of $L_{\rm max}$)
and/or the inclination has been seriously {\em overestimated} (lower
inclinations give a larger black hole mass and a larger value of
$L_{\rm Edd}$), the conclusion that $L_{\rm max} > L_{\rm Edd}$ seems
unavoidable.  Note that $L_{\rm max}$ was derived assuming the X-rays
were emitted isotropically.  If the X-ray emission was not isotropic,
then possibly the accretion rate during the 1983 outburst could have
been sub-Eddington.  However, detailed model computations should be
done, and such computations are beyond the scope of this paper.

\subsection{Accretion Stability Considerations}

As we discussed above, the temperature and radius of the secondary
star are consistent with a $2.5\,M_{\sun}$ star $\approx 393$~Myr past
the ZAMS.  Mass transfer caused by expansion while on the main
sequence is the so-called ``Case A'' phase (Kippenhahn \& Weigert
1967\markcite{kipp}).  If 4U 1543-47 does indeed contain a main
sequence A-type secondary the mass transfer will be stable since a
star with a radiative envelope will shrink in response to mass loss
(e.g.\ Soberman, Phinney, \& van den Heuvel 1997\markcite{sober}).  In
general, a low-mass X-ray binary will be a transient system if its
mass transfer rate is below some critical value (e.g.\ van Paradijs
1996\markcite{vanP96}).  The 1983 outburst released about $5.7\times
10^{45}$~ergs in X-rays (Chen, Schrader, \& Livio
1997\markcite{chen97} and cited references), assuming $d=9.0$~kpc.
For an average outburst interval of 10.5 yr, this corresponds to an
average X-ray luminosity of $\bar{L_x}=1.7\times
10^{37}$~ergs~s$^{-1}$.  The corresponding average mass transfer rate
(assuming an efficiency of $0.2mc^2$ per gram of accreted matter; van
Paradijs 1996\markcite{vanP96}) is $9.54\times 10^{16}$~g~s$^{-1}$
$=1.51\times 10^{-9}\,M_{\sun}$~yr$^{-1}$.  Based on his model of the
effects of X-ray irradiation on accretion disks, van Paradijs
(1996)\markcite{vanP96} gives the dividing line between transient and
persistent systems in the $P_{\rm orb} \,{\em vs.}\,\bar{L_x}$ plane
as $\log\bar{L_x} =35.8+1.07\log P_{\rm orb}$, where the orbital
period is in hours.  Systems below this line in the $P_{\rm orb}\,{\em
vs.}\, \bar{L_x}$ plane are transient.  With $P_{\rm orb}=26.95$ hours
and $\bar{L_x}=1.7\times 10^{37}$~ergs~s$^{-1}$, 4U 1543-47 is just
below the dividing line between transients and persistent sources.
Thus based on this quick analysis we conclude a main sequence
secondary can produce stable mass transfer in a transient system.

\section{Summary}

We have measured the radial velocity curve of the secondary star in 4U
1543-47.  The orbital period is $P=1.123\pm 0.008$ days and the $K_2$
velocity is $K_2=124\pm 4$~km~s$^{-1}$, which together give a mass
function of $f(M)=0.22\pm 0.02\,M_{\sun}$.  The spectral type of the
secondary is A2, and we see no emission lines from the accretion disk
and only a modest continuum flux from the disk in the $B$ and $V$
bands.  The $V$ and $I$ light curves are ellipsoidal, which implies
the A-star is the mass donor star and not a member of a triple system.
Furthermore, the spectroscopic and photometric periods agree, and the
relative phase of the radial velocity curve and the light curve is as
expected.  A model fit to the $V$ and $I$ light curves gives $24\le
i\le 36\arcdeg$ ($3\sigma$), with a lower limit of $i\ge 20\arcdeg$
and an upper limit of $i\le 40\arcdeg$.  The inferred radius, density,
and temperature of the secondary are consistent with those of a
$\approx 2.5\,M_{\sun}$ main-sequence star which is $\approx 393$~Myr
past the ZAMS.  If the secondary star is on the main sequence, then
the primary mass is in the range $2.7\le M_1\le 7.5\,M_{\sun}$.  We
conclude the compact object in 4U 1543-47 is very probably a black
hole.  Finally, we derive a distance of $d=9.1\pm 1.1$~kpc and a peak
X-ray luminosity during the 1983 outburst of $L_{\rm max}=3.6\times
10^{39}$ ergs s$^{-1}$.  The conclusion that the maximum X-ray
luminosity of the 1983 outburst exceeded the Eddington luminosity is
unavoidable ($3.1\lesssim L_{\rm max}/L_{\rm Edd} \lesssim 8.7$).

\acknowledgements

We thank George Jacoby for his donation of data, Richard Wade for his
help with the synthetic spectra, Andrew King for useful discussions,
and W. Niel Brandt for assistance with the ASCA archive.  We are
grateful to the staff of CTIO for the excellent support, in particular
Srs.\ Mauricio Navarrete, Edgardo Cosgrove, Mauricio Fernandez, Manuel
Hernandez, Patricio Ugarte, Hernan Tirdaro, \& Daniel Maturana.
Partial financial support for this work was provided by the National
Science Foundation through a National Young Investigator grant to
C. Bailyn, by the Smithsonian Institution Scholarly Studies Program to
J.  McClintock, and through a NASA funded grant administered by the
American Astronomical Society to J. Orosz.

\end{document}